\documentclass{sig-alternate-per}

\usepackage{lipsum}
\usepackage{url}
\usepackage{xcolor}
\usepackage{algorithm}
\usepackage{algpseudocode}
\usepackage{enumitem}

\AtBeginDocument{%
  }

\begin{document}

\title{CARINA: Carbon-Aware Execution of Recurrent Industrial Analytics}

\CopyrightYear{2026} 

\author{
Muhammad Umar Farooq\\
\affaddr{Department of Mechanical Engineering, University of Michigan}\\
\affaddr{Ann Arbor, MI, 48109, USA}\\
\affaddr{muf@umich.edu}}

% The "conferenceinfo" command is used to define the location and workshop information
\conferenceinfo{Workshop on Measurements, Modeling, and Metrics for Carbon-Aware Computing (CarbonMetrics) 2026}{Ann Arbor, MI, USA}

\maketitle

\begin{abstract}
Recurring industrial analytics and machine-learning workflows are becoming a major computational burden in modern engineering practice. Large parametric database generation, scheduled model retraining, repeated evaluation pipelines, and extensive hyperparameter exploration can demand hundreds of runtime hours and tens of kilowatt-hours per refresh cycle, yet these workloads are rarely executed with explicit energy-awareness. We present CARINA (Carbon-Aware Recurrent Industrial Analytics), a measurement-and-estimation framework for energy-aware and carbon-aware execution of recurrent analytics. The framework combines lightweight run-level and step-level instrumentation, peak-time-aware execution control, and local dashboard reporting. The method estimates energy load as the primary objective and translates it to carbon emissions using a local grid emission factor, enabling use even when direct device-level carbon metrology is unavailable. We evaluate the framework using two automotive OEM database-generation workflows. The first required 1.48 million scenarios, 180.30 h, and 48.67 kWh; the second required 3.66 million scenarios, 274.75 h, and 74.16 kWh (corresponding to approximately 21.8 kg CO$_2\text{e}$ and 33.2 kg CO$_2\text{e}$, respectively). Preliminary policy analysis suggests that peak-aware off-hours boosting can reduce full-cycle energy load by about 9\% with roughly 7\% runtime overhead, while naive throttling can increase total energy through overhead effects.

\end{abstract}

\section{Introduction}

Industrial decision-making increasingly depends on computational workflows that are executed repeatedly rather than once \cite{bouza2023estimate}. These include large parametric database generation, design-space exploration, scheduled data-processing pipelines, periodic model retraining, validation campaigns, and extensive hyperparameter search \cite{kinkar2022carbon}. Their significance lies not only in individual run cost, but in recurrence: each refresh cycle, retraining round, or database rebuild contributes to a persistent operational energy demand. Wall-clock runtime is routinely reported in engineering simulation, yet comparable reporting is far less common for other industrial analytics and recurring machine-learning workflows \cite{santosh2026ai}, leaving their execution burden largely invisible until it has accumulated.

Recent literature has shown that the environmental footprint of computation depends on workload scale, hardware choice, runtime, memory allocation, and electricity context. Prior work in machine learning has emphasized the energy and emissions implications of modern training workloads, while subsequent tools and frameworks have made monitoring, prediction, and standardized accounting more practical \cite{bouza2023estimate, khan2025optimizing}. For example, CarbonTracker \cite{anthony2020carbontracker} and related efforts \cite{wegmeth2025green} showed that training-time monitoring and prediction can make such reporting practical. In parallel, Green Algorithms \cite{lannelongue2021green} provided a standardized framework for estimating the carbon footprint of computation from runtime, hardware, memory, and infrastructure assumptions, and tools such as CodeCarbon \cite{lottick2019energy} translated these ideas into lightweight local instrumentation, while many \cite{mane2025balancing} extended dashboard-style accounting to scheduler-managed environments. 

However, recurrent industrial analytics remains insufficiently addressed. Existing work is strongest either at retrospective accounting or at platform-level monitoring, whereas many real industrial workloads still execute on engineering workstations or lightly managed servers and require execution-time control as well as reporting. In these settings, users need to know not only how much energy a workflow consumed, but also how execution intensity should be adapted across the day so that recurring computations remain both energy-aware and operationally practical. This need spans both industrial database-generation workflows and repeated machine-learning training, testing, and evaluation pipelines.

This paper addresses that gap through \textbf{CARINA} (\textbf{C}arbon-\textbf{A}ware \textbf{R}ecurrent \textbf{IN}dustrial \textbf{A}nalytics), a estimation-based rather than device-meter-based framework for the execution of recurrent industrial analytics. CARINA combines lightweight instrumentation, system auto-detection, peak-time-aware worker control, and local dashboard reporting so that users can observe energy burden, translate it into local carbon burden, and compare execution policies with minimal change to existing code.

\section{Approach}

CARINA is designed for recurrent industrial analytics. It treats a workload as a sequence of tracked units, where a unit may be a full run, a refresh batch, a wave, an epoch, or a training round. For each unit, CARINA records runtime, selected worker intensity, estimated energy load, translated carbon burden, and execution metadata. The framework then aggregates these unit-level records into a run summary and compares execution policies against a measured baseline.

The control logic is based on fixed clock-time execution policies. Instead of running at a constant intensity throughout the day, CARINA adapts worker intensity according to local time bands such as peak, load-sensitive, shoulder, and night periods. In practice, the framework reduces worker intensity during sensitive periods and increases it later to recover throughput. The current implementation also supports system auto-detection, thread capping, CPU affinity restriction, priority reduction during load-sensitive hours, structured logs, and dashboard generation. 

\begin{algorithm}
\caption{CARINA Execution and Tracking}
\begin{algorithmic}[1]
\State \textbf{Input:} Workload $W$, timing policy $P$, machine profile $M$, local grid factor $g$
\State \textbf{Output:} Energy-aware and carbon-aware execution record
\State Detect machine characteristics and initialize tracker
\State Choose tracking granularity: \textit{whole-run} or \textit{step-level}
\For{each tracked unit $W_i \in W$}
    \State Determine the local time phase
    \State Select worker intensity from policy $P$
    \State Apply thread / priority / affinity controls
    \State Execute $W_i$
    \State Record runtime, energy estimate, and execution metadata
    \State Translate energy to carbon using the local grid factor
\EndFor
\State Aggregate totals across tracked units
\State Save logs, summary metrics, plots, and dashboard artifacts
\State Compare policy outputs against the measured baseline
\State \Return Execution record and analytics summary
\end{algorithmic}
\end{algorithm}

We validate CARINA on two sheet-metal database-generation workloads from an automotive OEM context. Both are built on an Excel-native parametric analytics model with a Python add-on for large-scale offline scenario generation. The design spaces span roughly ten key variables per case and combine categorical and numerical choices such as feedstock composition, process yields, alloy or route selection, and regional electricity mixes. Scenarios are designed in Python and executed against worker-local Excel copies using batched evaluation, where each batch writes inputs, triggers recalculation, extracts outputs, and stores results. The implementation uses batched parallel execution together with resume, merge, and verification logic, making the workloads a realistic validation target for CARINA’s tracking and control framework.

\section{Results and Discussion}

\begin{figure}[htbp]
    \centering
    \includegraphics[width=\linewidth]{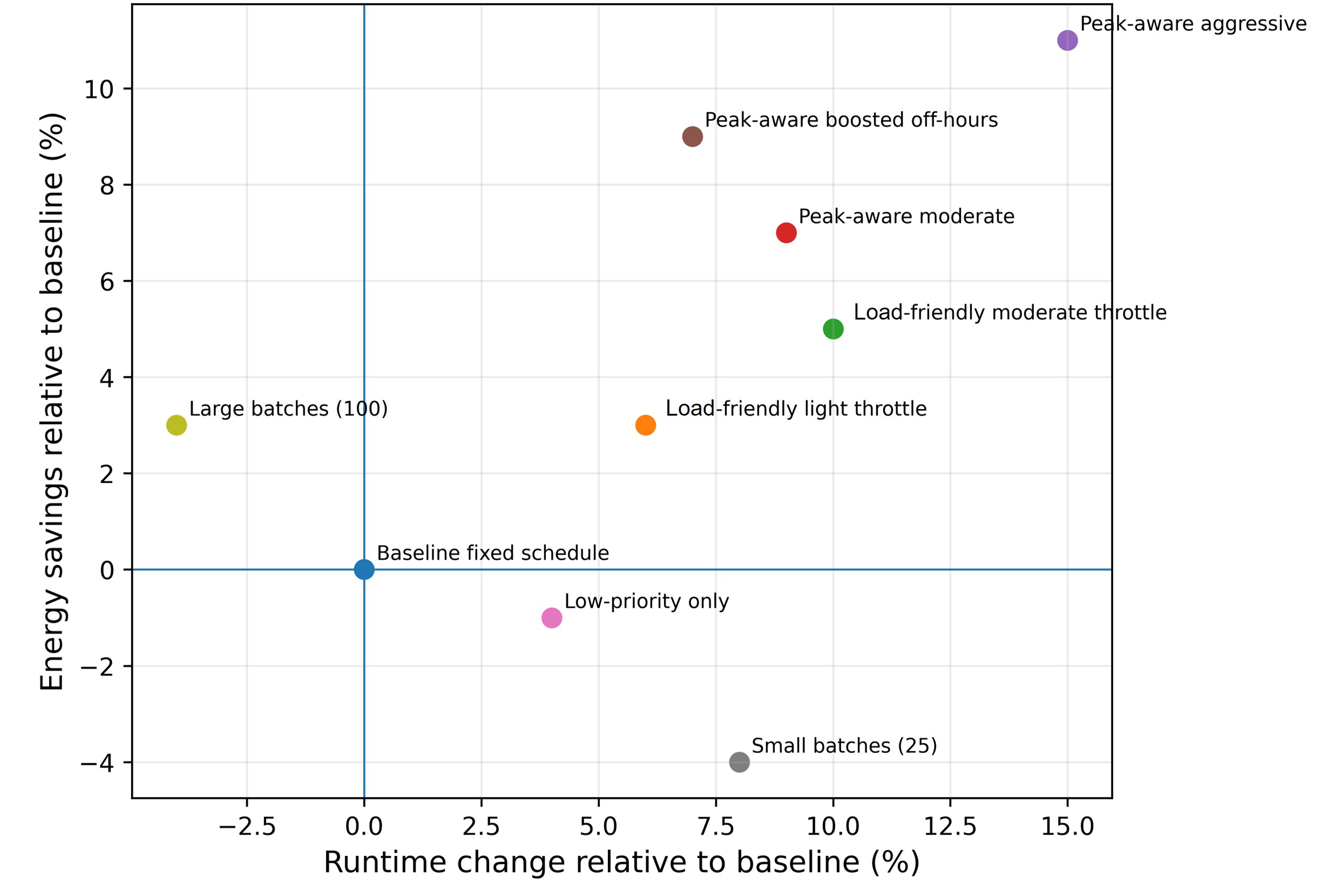}
    \caption{Modeled trade-off between runtime penalty and energy savings for candidate full-refresh execution policies relative to the measured baseline.}
    \label{fig:tradeoff}
\end{figure}

Figure~\ref{fig:tradeoff} presents a normalized comparison for recurrent large-scale computational workloads. The average numbers are based on the automotive OEM database-generation demonstrations, and demonstrated recurrent ML workloads such as scheduled federated learning (PACS Vision Dataset), tabular-model retraining (Critical Gas Power Plant Operations), and large hyperparameter search campaigns (Laser Powder Bed Fusion of Commercial Alloys). It summarizes how execution policies trade runtime overhead against energy-load savings relative to the measured baseline.

Firstly, peak-aware control moves the workload into the most favorable region of the frontier. \textit{Peak-aware boosted off-hours} provides the best overall compromise, with modeled energy-load savings of about \textbf{9\%} for roughly \textbf{7\%} runtime overhead, while \textit{peak-aware aggressive} delivers the largest savings at the highest completion-time cost. Second, not all interventions are beneficial: \textit{low-priority only} slightly increases total energy use, and \textit{small batches (25)} performs worse still because orchestration overhead grows faster than the benefit of throttling. Third, \textit{large batches (100)} improve both runtime and energy, indicating that startup and coordination overhead are major determinants of total burden.

The two OEM case studies show the practical scale of this problem. The first workflow consumed \textbf{48.67 kWh}, corresponding to approximately \textbf{21.8 kg CO$_2$e} under the Detroit-area DTE factor, while the second consumed \textbf{74.16 kWh}, or approximately \textbf{33.2 kg CO$_2$e}. Applying the most balanced peak-aware policy in the same proportion as Figure~\ref{fig:tradeoff} reduced these workloads to approximately \textbf{44.3 kWh} and \textbf{67.5 kWh}, corresponding to savings of about \textbf{4.4 kWh} (\textbf{2.0 kg CO$_2$e}) and \textbf{6.7 kWh} (\textbf{3.0 kg CO$_2$e}), respectively, at an estimated runtime increase of about \textbf{7\%}.

Overall, the policy chart and the measured OEM baselines show that recurrent industrial analytics is already large enough to justify execution-aware energy management. The main value of CARINA is therefore not only to report burden, but to distinguish effective controls from counterproductive ones. Policies that shift execution away from peak periods can reduce total burden, whereas naive throttling or poorly chosen batch structure can increase it through overhead effects.

\section{Conclusions}
Recurrent industrial analytics and recurring ML workflows constitute substantial computational burdens whose energy and carbon implications should be managed explicitly. CARINA shows that peak-aware execution control can reduce total burden meaningfully, while some intuitive interventions can instead increase energy use through overhead effects. At present, the framework relies on energy-load estimation and local factor translation rather than direct physical power measurement or continuously updated regional carbon-intensity feeds. Future work should therefore combine CARINA with hardware-based validation and, time-varying local grid carbon data.

\bibliographystyle{unsrt} %abbrv
\bibliography{paper}

% \clearpage
% \section*{Outline}
% \input{00-outline}

\end{document}